\documentstyle[11pt,epsf]{article}
\newcommand{\be}{\begin{equation}}
\newcommand{\ee}{\end{equation}}
\newcommand{\beq}{\begin{eqnarray}}
\newcommand{\eeq}{\end{eqnarray}}

\begin{document}
\thispagestyle{empty}


\title{The Deconfinement Phase Transition in One-Flavour QCD}

\author{Constantia Alexandrou$^{\rm a}$,  Artan Bori\c{c}i$^{\rm b}$, 
Alessandra Feo$^{\rm a}$,\\
Philippe de Forcrand$^{\rm c}$, Andrea Galli$^{\rm d}$, 
Fred Jegerlehner$^{\rm e}$, Tetsuya Takaishi$^{\rm f}$\\ 
\\
$^{\rm a}${\it Department of Natural Sciences, 
        University of Cyprus, CY-1678 Nicosia, Cyprus}\\
$^{\rm b}${\it Paul Scherrer Institute,
        CH-5232 Villigen PSI, Switzerland}\\
$^{\rm c}$ {\it Swiss Center for Scientific Computing, 
        ETH-Zentrum, CH-8092 Z\"urich, Switzerland}\\
$^{\rm d}${\it ELCA Informatique, HofwiesenStr. 26,          
        CH-8057 Z\"urich, Switzerland}\\
$^{\rm e}${\it DESY-IfH Zeuthen, D-15738 Zeuthen, Germany}\\
$^{\rm f}${\it Hiroshima University of Economics, Hiroshima, Japan 731-01}
\\
UCY - PHY - 98/05\\
}
\maketitle

\begin{abstract}
We present a study
of the deconfinement phase transition  of one-flavour QCD, 
using the multiboson algorithm. The mass of the Wilson fermions
relevant for this study is  moderately large and  
the non-hermitian multiboson method is a superior simulation 
algorithm.
Finite size scaling  is studied on lattices of size 
$8^3\times 4$, $12^3\times 4$ and $16^3\times 4$. The behaviours of 
the peak of the Polyakov loop susceptibility, 
the deconfinement ratio and the distribution of the norm of
the Polyakov loop 
are all characteristic of  a first-order phase
transition for heavy quarks. 
As the quark mass decreases,
the first-order transition gets weaker and turns into
 a  crossover.
To investigate finite size
 scaling on larger spatial lattices we use
 an effective action in the same universality class as QCD.
This effective action is constructed by 
replacing  the fermionic determinant
 with the Polyakov loop identified as  the most relevant  $Z(3)$ 
symmetry breaking  term.
Higher-order
effects are incorporated in an effective $Z(3)$-breaking field, $h$,
which couples to the Polyakov loop. Finite size scaling determines the value
of $h$ where the first order transition ends. Our analysis at the end - point,
$h_{ep}$, indicates that the effective model and thus QCD 
is consistent with 
the universality class of the three dimensional Ising model.  
  Matching the field
strength at the end point, $h_{ep}$, 
to the $\kappa$ values 
used in the dynamical quark simulations
we estimate the end point, $\kappa_{ep}$, of the first-order phase
transition. We find $\kappa_{ep}\sim 0.08 $ which corresponds to 
  a quark mass of about 1.4 GeV . 



\end{abstract}

\section{Introduction}
Understanding the properties of Quantum Chromodynamics (QCD) under extreme
conditions of high temperature and/or pressure is a challenging problem
and was considered long ago.
 Polyakov \cite{Polyakov} and Susskind \cite{Suss}
predicted that QCD will undergo a deconfinement phase transition from 
normal hadronic matter to quark-gluon plasma when the temperature is increased.
Such a phase transition is believed to have occurred,
 in the opposite direction, $1 {\mu s }$ after the Big Bang and its nature 
 is therefore important in Astrophysics.
Theoretical information on the deconfinement phase transition   
has also become  important
as planned ultra-relativistic experiments will soon start at RHIC, Brookhaven
and later on at LHC, CERN.
In these experiments the temperature reached will be of the order of 600 MeV 
and, at this temperature, one is still dealing with a strongly 
interactive system.
Thus Lattice QCD provides the most suitable non-perturbative approach 
to study such phenomena using directly the QCD Lagrangian.

The zero flavour sector of the theory (quenched) has been studied extensively
and it is established that there is a first - order
deconfinement  phase transition \cite{Fuk} with the critical temperature 
determined in the continuum limit~\cite{Iwas,Edw,Boyd,DeGrand,Beinl}.
State of the art lattice calculations are now being done including pair 
creation. For two flavours ($N_f=2$) there  are simulations both with Wilson
\cite{MILC} and staggered fermions \cite{MILC,lat97} whereas for greater
 number
of flavours the simulations are usually performed with the standard Wilson action 
\cite{lat97,Iwas2}. On the other hand, 
  one - flavour QCD has been largely ignored (early exceptions
are given in ref. \cite{nf1}), perhaps because of algorithm difficulties,
 although it has interesting properties.
 The continuum $N_f=1$ theory contains no pions and it is  expected to have 
no chiral phase transition\cite{PW} with the results
 of a study of 
one flavour staggered fermion  
\cite{ChNorm}  implying a  
chiral phase transition coming as a surprise.
The purpose of this work is
to fill this gap by investigating the deconfinement phase transition 
of  one - flavour QCD. 

\begin{figure}[hbtp]
\begin{center}
\mbox{\epsfxsize=7cm\epsfysize=5cm\epsffile{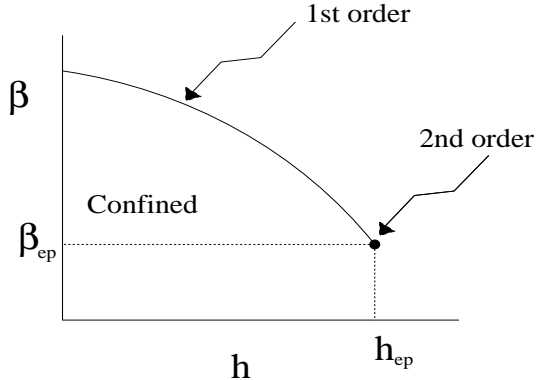}}
\end{center}
\caption{ Phase diagram in the three states Potts model in three dimensions}
\label{fig:bc_diag}
\end{figure}

A model that
plays an important role in our understanding of the phase diagram  in 
QCD is the three states Potts model in three dimensions.
 It has a Z(3) symmetry and the
spontaneous break down of this symmetry is expected to drive the deconfinement
phase transition  like it does in quenched QCD. 
 In the presence of an external field the  Z(3)  symmetry  is explicitly
broken just like the fermionic determinant breaks Z(3) symmetry
in QCD.  In fact it was shown by DeGrand and DeTar~\cite{DeGDeT}
 that, at high temperature
and heavy quark mass, QCD reduces to the 3 - states Potts model in
an external field. From its phase diagram~\cite{DeGDeT},
 shown schematically  in Fig.~\ref{fig:bc_diag}, 
  we observe  that the first order
phase transition gets weaker as the
strength of the external field $h$ increases and ends at some critical value
$h_{ep}$ with a second order transition. Beyond this point a crossover 
behaviour is seen. 
Whereas Fig.~\ref{fig:bc_diag} gives us a good starting point
 for the expected qualitative behaviour for QCD with  dynamical quarks, 
the quantitative question of existence of such an end-point
 and its 
location can only be answered after a detailed
calculation.

In the real world of two light and one heavy quarks the
phase diagram depends crucially on the size of the light and heavy 
quark masses. Therefore it is usual to consider the phase diagram of 3-flavour
QCD in the mass plane $m_s$ vs $m_{u,d}$. The mass point (0,0) is the chiral
limit with a first order  transition.
Pure gauge is the  opposite limit of infinite quark masses where  a first
order transition is also established. 
\noindent
 As one moves away from the infinite mass (or hopping parameter $\kappa=0$) limit
the first order transition is expected, like in the Potts model, 
to persist but to  become weaker
and eventually to
disappear, as the quark mass is reduced.
 This presumed robustness is based on the expectation 
 that an infinitesimal increase
in $\kappa$  does not cause the finite gap in the
thermodynamical observables to change discontinuously 
but smoothly. That a first order transition is in general robust
 is supported from our experience with   spin systems 
where a first order transition remains
first order for small magnetic fields. 
In this work we seek   to determine the point on the line 
$(m_{u,d}=\infty,m_s)$ where the first order transition ends as well
as the universality class of the second order phase transition
at this end point.
 It turns out that the relevant quark
masses for this study are relatively large and therefore the multiboson
algorithm is a very efficient method for simulating such dynamical quarks.

From the determination of the end point for one - flavour 
we can draw conclusions about the
end point for two flavours i.e. where on the line $(m_{u,d},m_s=\infty)$ 
the first order transition ends. 
 The quark mass where this is expected to occur is
about hal the mass obtained in the  one - flavour case  because
 the quark effects that we observe are about half as large as the effects
obtained for two flavours  of the same mass.

In order to determine quantitatively the phase diagram for QCD 
we perform simulations on lattices with size 
$8^3 \times 4$, $12^3 \times 4$ and $16^3 \times 4$.
Performing a finite size scaling analysis
 we find  indeed that
there is a first order phase transition for heavy quarks which gets weaker as
the quark mass decreases and
becomes  second order at a value of $\kappa_{ep} \sim 0.08$. For smaller
quark masses a crossover behaviour is seen~\cite{lat}. 

In order to determine the end point in the ($\beta,\kappa$) plane
accurately one needs larger lattice sizes, which  takes
 very long to simulate. However there is a different route that we can follow.
 Namely we  investigate an effective model
in the same universality class as QCD which is
easier to simulate, determine the end point there and then match back to QCD.  
Such an effective model can be constructed by taking the 
pure gauge action and adding the most relevant Z(3) breaking term
which is a Polyakov loop. The strength of the Z(3) breaking
field is adjusted to model  higher order terms.
 Within this effective 
model we were able to perform a finite - size scaling analysis  up to a lattice
size of  $24^3 \time 4$ and determine the critical strength of the Z(3)
breaking field where the first order transition ends. We also 
investigated the lattice spacing dependence within the 
effective model, by performing
a simulation for temporal extension $N_t=2$ in addition to $N_t=4$.
Finite size scaling at the end - point,
$h_{ep}$, yields results that are consistent with the scaling behaviour 
seen in the three dimensional Ising model. 
We obtain $h(\kappa)$ by performing a best fit of the Polyakov loop histograms
obtained in the effective model to those obtained in QCD.
This non perturbative matching of $h_{ep}$
yields the critical
value of $\kappa_{ep}$ and of $\beta_{ep}$.  

This paper is organized as follows: In section (2) we  give details on the
local bosonic algorithm which we used to simulate one dynamical Wilson fermion.
In section (3) we describe the observables that we used to probe  a
change of phase  and the nature of this phase transition. We give
the results of this analysis in section (4). In section (5)
we discuss the simulation and the finite size scaling
analysis of the effective model and connect it   to
 full QCD. In section (6) we discuss the continuum limit and
finally in section (7) we summarize and  conclude.

\setcounter{equation}{0}

\section{Local bosonic algorithm for one flavour}
\begin{figure}
\begin{center}
\mbox{\epsfxsize=8cm\epsfysize=7.5cm\epsffile{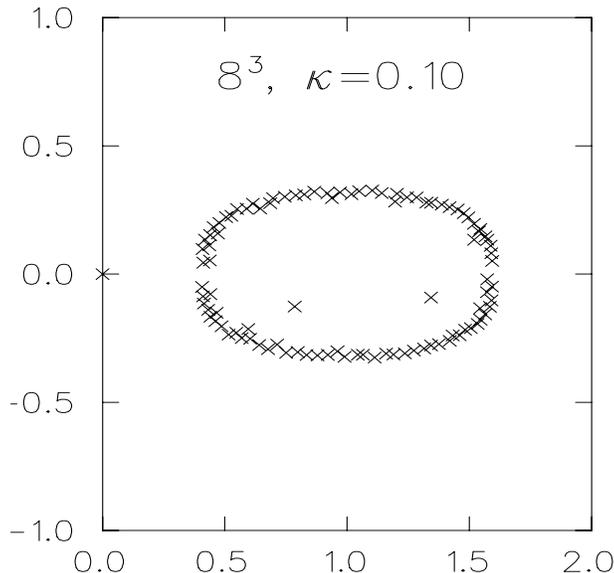}}
\end{center}
\caption{ Estimated boundary of the Dirac eigenvalue spectrum}
\label{fig:eig_plot}
\end{figure}

The local bosonic algorithm was originally proposed by L\"uscher \cite{Luscher}
as an alternative method to the widely used Hybrid Monte Carlo (HMC) algorithm 
to simulate dynamical quarks. The basic idea is the
replacement of the fermionic determinant by a functional integral over 
$n$ bosonic fields having a local action. If $Q=\gamma_5 D/(1+8\kappa)$
where $D$ is the fermionic Wilson matrix then for two degenerate flavours
we have 
\be
det Q^2 \propto \lim_{n\rightarrow \infty} 
          \int\prod_{k=1}^{n} d\phi^{\dagger}_k d\phi_k 
         e^{-\sum_{k=1}^{n} \phi^{\dagger}_k 
                         \left [ (Q-\mu_k)^2 + \nu_k^2 \right] \phi_k}
\label{det Q2}
\ee
where $\sqrt{z_k}=\mu_k + i \nu_k$ with $z_k$ being the roots of a polynomial
of even degree $n$ constructed so that 
\be
\lim_{n\rightarrow \infty} P_n(Q^2)=\frac{1}{Q^2} \quad.
\label{P_n}
\ee
The error introduced by taking $n$ finite can be eliminated by a global 
accept/reject Metropolis test \cite{BFG}. 
The generalization to any number of flavours is made possible by finding
a polynomial approximation to the fermionic matrix itself rather than to
$Q^2$ \cite{BFG,BF}.  
This can be done by constructing a polynomial of even degree $n$  
defined in the complex plane with complex conjugate roots $z_k$ 
such that
\be
\lim_{n\rightarrow \infty} P_n(z)= \frac{1}{z}
\label{P_n complex}
\ee
for any $z$ in the  domain containing the spectrum of $D$ 
(not including the origin). 
Since the spectral radius of the hopping matrix $M$ is bounded by 8
in the free case and  even less in the interacting case, 
we are guaranteed that
the spectrum of $D = 1 - \kappa M$ will remain in the complex right half-plane
for the heavy to moderately heavy quarks that we simulate ($\kappa \leq 1/8$).
This is demonstrated in Fig.~(\ref{fig:eig_plot}) where the boundary of 
the spectrum of the Dirac matrix  is estimated for $\kappa=0.1$
as in ref.~\cite{lat98}. 

Using the property $D=\gamma_5 D^{\dagger} \gamma_5$ we obtain
\be
det P_n(D)=c_n \prod_{k=1}^{n/2} det(D-z_k)^{\dagger}det(D-z_k) 
\label{det P_n }
\ee
with $c_n$ an easily computed  constant \cite{BF}.
Instead of eq. (\ref{det Q2}) one now finds
\beq
det D &=& \lim_{n\rightarrow \infty} det^{-1} \left [ T^{\dagger}_{n/2}(D)
                                         T_{n/2}(D)\right] \\ \nonumber  
 &\propto & \lim_{n\rightarrow \infty} 
          \int\prod_{k=1}^{n/2} d\phi^{\dagger}_k d\phi_k 
         e^{-\sum_{k=1}^{n/2} \phi^{\dagger}_k 
                          (D-z_k)^{\dagger}(D-z_k) \phi_k} \quad ,
\label{det D}
\eeq 
where $T_{n/2}(D)=\prod_{k=1}^{n/2} (D-z_k)$.
The algorithm is made exact with a global Metropolis test as follows.
The one-flavour determinant can be expressed as
\be
{\rm det} D = {\cal C}{\rm det}^{-1} \left [T^{\dagger}_{n/2}(D) T_{n/2}(D)\right]
\label{correction}
\ee
with the correction factor ${\cal C}$ given by
\be
{\cal C} = \lim_{m\rightarrow \infty} 
          \frac{{\rm det}^{-1}\left [T^{\dagger}_{m/2}(D) T_{m/2}(D) \right]}
               {{\rm det}^{-1}\left [T^{\dagger}_{n/2}(D) T_{n/2}(D) \right]}
\label{correction term}
\ee
Because the error of the polynomial approximation (\ref{P_n complex})
decreases exponentially with the degree of the polynomial, it is not 
necessary to actually take the limit $m\rightarrow \infty$ above.
We observed that taking $m \geq 3 n$, where $n$ is adjusted for sufficient
Metropolis acceptance below, is enough to make the systematic approximation 
error completely negligible compared to the statistical one.
We implemented the correction factor ${\cal C}$ by a noisy Metropolis test,
along the lines suggested in ref.\cite{BFG}.
The gauge and boson fields $(U, \phi)$ are updated by a sequence of local
Monte Carlo steps forming a trajectory $(U, \phi) \rightarrow (U',\phi')$.
This procedure satisfies detailed balance with respect to the approximate
action 
\be
S_{\rm approx} = S_g[U] + \sum_{k=1}^{n/2} |(D-z_k) \phi_k|^2
\label{S_approx}
\ee
At the end of each trajectory, one generates 
a field $\eta$ with probability
\be
 P^{HB}(\eta) = \>R e^{-|X\eta|^2}
\ee
where $X=T_{m/2}(D)T^{-1}_{n/2}(D)$ and $R$ is a normalization constant.
The new configuration 
$(U',\phi')$ is accepted with probability 
\be 
P^{A}_{(U,\phi)\rightarrow (U',\phi')}
  ={\rm min}\left[1, \frac{e^{-|X'\eta|^2}}{e^{-|X\eta|^2}} \right] \quad.
\ee
The correction term is thus estimated by only one $\eta$ field. 
Detailed balance with respect to the desired action
${\rm det}^{-1}\left [T^{\dagger}_{m/2}(D) T_{m/2}(D) \right] e^{-S_g[U]}$ 
is satisfied after averaging the probability density for the Metropolis test, 
$P^{HB}(\eta)P^{A}_{(U,\phi)\rightarrow (U',\phi')}$, over the $\eta$ field.

The number of bosonic fields $n$ is chosen so that the correction term
leads to an acceptance  rate of about 2/3  and $m$ is taken
at least three times $n$. 
  
For the local updating of the gauge and boson fields we used standard heatbath
and over-relaxation algorithms as described in \cite{BFG}.
A trajectory 
is a symmetric combination of $(2 l + 1)$ over-relaxation steps applied 
alternatively to the gauge and boson fields, preceded and followed by
a heatbath on the bosons.
Ergodicity for the gauge fields is
maintained due to their coupling to the bosonic fields. 
 The roots $z_k$ are
distributed on a circle centered at (1,0). We implemented
 even - odd preconditioning 
 to lower the number of bosonic fields needed for a given accuracy. 
To efficiently equilibrate the system,
we start from thermalized quenched gauge configurations  and initialize
the boson fields by generating a gaussian random vector $\chi$ and setting
\be
\phi_k \leftarrow c_n~D \prod_{l\neq k}^{n} (D-z_k)~\chi \quad.
\label{quasi heatbath}
\ee

\setcounter{equation}{0}

\section{Signals for the phase transition or lack thereof}

QCD with infinitely heavy quarks, i.e. quenched, undergoes a first-order
deconfinement transition corresponding to the spontaneous breaking of the
$Z(3)$ Polyakov loop symmetry~\cite{Sve}.
 Based on the Polyakov loop, given by  the
product of gauge links in the temporal direction,
\be
L({\bf n})=\frac{1}{3} {\rm Tr}\prod_{n_0=1}^{N_t} U_0({\bf n},n_0) \quad ,
\label{Polyakov loop}
\ee
which transforms under $Z(3)$ as  $L({\bf n}) \rightarrow zL(\bf{n})$,
one can construct a standard set of observables and order parameters.
In particular, $\langle L \rangle = 0$ in the $Z(3)$-symmetric, confined
phase, and non-zero in the spontaneously broken, deconfined phase.

As explained in the Introduction,
one expects this phase transition to persist in the presence of sufficiently
heavy dynamical quarks. However, these dynamical quarks explicitly break
the quenched $Z(3)$ symmetry, favoring the real $Z_3$ branch and inducing
a non-zero Polyakov loop $\langle L \rangle > 0$ even in the confined phase.
The Polyakov loop and associated observables can no longer serve as order
parameter. The phase transition will be identified as a discontinuity in 
these observables, from one non-zero value to another, in the thermodynamic
limit.

Furthermore, 
as also explained in the Introduction,
this first-order phase transition is expected to become weaker and eventually
disappear, as the quark mass is reduced. The correlation length at criticality
will correspondingly increase, and eventually diverge at the end-point of
the transition line where the transition becomes second-order. For yet lighter
quarks, no singularity appears even in the thermodynamic limit, and a simple 
crossover occurs in $\langle L \rangle$. The distinction between weak 
first-order, second-order, or crossover behaviour requires lattice sizes 
at least comparable with the critical correlation length. However, this
requirement can be lessened if one compares results for various small size
lattices with a finite-size scaling ansatz. This is how the first-order nature
of the quenched transition was first ascertained \cite{Fuk}, and this is how
we proceed here.

Our strategy is to vary $\beta$, for each quark mass, 
for three spatial lattice sizes and
 to look for the following signals: 
\begin{itemize}
\item {\it Coexistence of the two phases:}
A distinctive feature of a first order transition is phase coexistence
and on a finite lattice  
 we look for tunneling between the
confined and the deconfined phase which is observed over a small temperature
range around the  critical temperature. As the size of the lattice increases
tunneling is exponentially
 suppressed but for the lattice sizes studied here enough 
tunneling events are observed to enable us to study the
  double peak distribution for the norm $|\Omega|$ of the Polyakov loop
 defined as 
\be
\Omega=\frac{1}{V} \sum_{\bf n}L(\bf{n}) 
\label{Omega}
\ee
with $V$ the spatial volume.
For a first-order phase transition, the location of the peaks should be fixed
as the volume increases, while their width should decrease like $V^{-1}$.
For a second-order phase transition or a crossover, the peaks should merge
as the volume increases, and the tunneling signal should disappear.

\item {\it Deconfinement ratio:} The deconfinement ratio is defined by 
\be
\rho = \frac{3}{2}\> p -\frac{1}{2}
\label{rho}
\ee
with $p$ the probability for the complex Polyakov loop to be within 20 degrees
of a $Z_3$ axis. Therefore if the $Z(3)$ symmetry is unbroken we find  $p=1/3$
and $\rho=0$, whereas if it is broken in such a way that the Polyakov loop is 
distributed near one axis (the real axis)  we have 
$p=1$ and $\rho=1$.   The value of $\rho=0$ 
is only obtained in the quenched case where the  $Z(3)$ symmetry is exact.
With dynamical quarks we only look for a discontinuity across the phase transition.
On a finite lattice, the discontinuity is smoothed out. $\langle \rho \rangle$
varies abruptly over a range in $\beta$ which shrinks as $V^{-1}$, and the slope
$d\rho / d\beta$ is maximum at the critical $\beta$.
The intersection of curves $\rho(\beta)$ for various lattice sizes may also
give information on $\beta_c$.
\footnote{ One complication here is that $\langle \rho \rangle$ in the confined
phase depends on the spatial volume $V$.}

\item {\it The real part of the Polyakov loop after projection on the
closest $Z_3$ axis:} This observable is also an order parameter for the quenched
phase transition. It behaves just like the deconfinement ratio in the presence
of dynamical quarks.

\item {\it The peak value of the susceptibility:}  The susceptibility measures
the fluctuations of the Polyakov loop.
We consider the behaviour of
\be
\chi_L=V(\langle |\Omega|^2 \rangle - \langle |\Omega| \rangle^2)
\label{chi}
\ee
which diverges at criticality for a first-order phase transition
in the thermodynamic limit.
On a finite lattice, instead of this $\delta$ function  behaviour,
the peak value of $\chi_L$ is proportional to
the volume $V$, while the width of the  $\chi_L$
distribution scales like $V^{-1}$ and its peak may shift 
like $V^{-1}$~\cite{ising}.
The scaling behaviour  of the susceptibility changes for a
 second order transition: the peak value of $\chi_L$
 becomes proportional to  $V^{\alpha}$ with $\alpha<1$. 
For a crossover behaviour where, even
 in the thermodynamic limit, there is
 no discontinuity in the 
thermodynamic functions, the peak value of  $\chi_L$ remains constant with
the volume.  
\end{itemize}

\section{Results}

\begin{figure}[hbtp]
\begin{center}
\mbox{\epsfxsize=14cm\epsfysize=9cm\epsffile{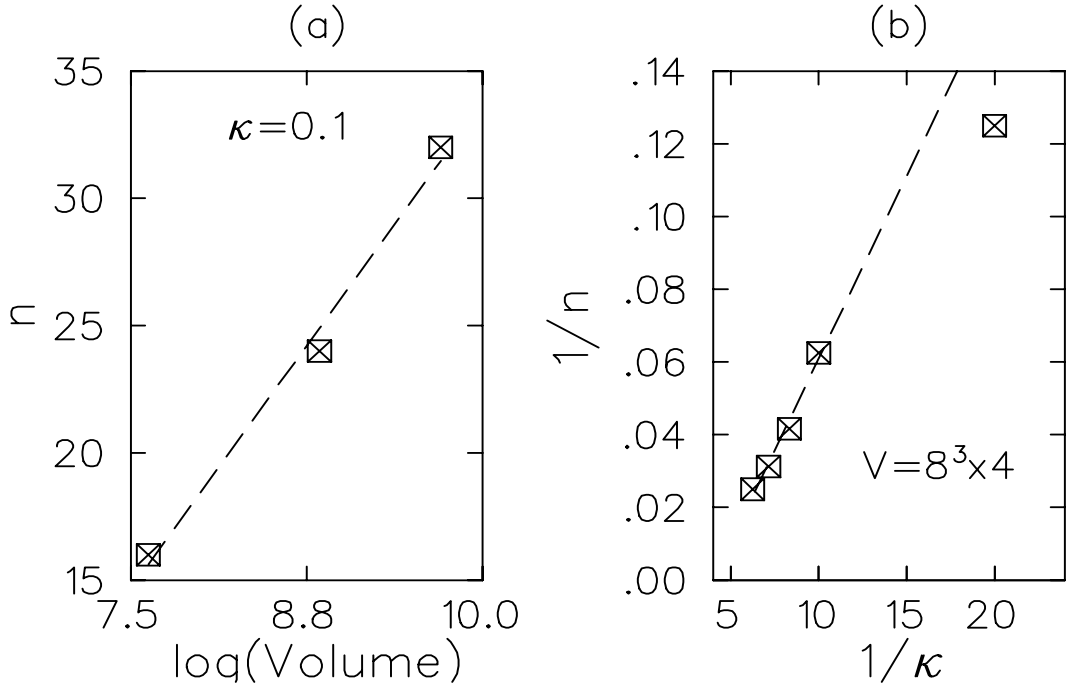}}
\end{center}
\caption{(a) The 
 number of bosonic fields, $n$, is plotted vs the logarithm of the volume
for $\kappa=0.1$. 
(b) 
 $1/n$ vs $1/\kappa$ for the $8^3\times 4$ lattice.}  
\label{fig:boson_param}
\end{figure}

In Table \ref{table:parameters} we collect the parameters of our simulations.
Adjusting the number of bosonic fields, $n$, 
 so that the acceptance is about 2/3, we can make two observations:
Firstly, for a fixed quark mass 
we find that $n$ grows logarithmically with the volume
for the same acceptance ratio (i.e. for the same 
relative error in the bosonic approximation of
the determinant). This is 
illustrated in   Fig.~\ref{fig:boson_param} which
shows the $\kappa=0.1$ data
from Table~\ref{table:parameters}.
Secondly, for a fixed volume, $n$ is approximately inversely
proportional to the quark mass i.e. $1/n$ is linear in $1/\kappa$. This 
behaviour is
displayed in Fig.~\ref{fig:boson_param} where it holds for a range of $\kappa$
values. 
This dependence of $n$ on both the quark mass and the volume confirms
the expectations of e.g. Ref.~\cite{lat95}.

\begin{table}[h]
\caption{\label{table:parameters}}
In the first column we give the $\kappa$ values for the three 
volumes studied. In the other columns $n$ is the number of bosonic
fields and $acc$ the average acceptance; $Ksw$ gives (in kilo sweeps) 
the total number 
of thermalized configurations used in the
reweighting procedure \cite{FeSwend}.
\begin{center}
\begin{tabular}{|c||cc|cc|cc|}
\hline
$\kappa$ &\multicolumn{2}{c}{ $8^3 \times 4$} &
    \multicolumn{2}{c}{$ 12^3 \times 4$} & 
\multicolumn{2}{c|}{$ 16^3 \times 4 $ }\\
\hline
 & $n/acc$ & $Ksw$ & $n/acc$ & $Ksw$ & $n/acc$ & $Ksw$ \\
\hline
 0.05 & 8/0.78 & 18 & 12/0.74 &20 & 24/0.83 & 20 \\
\hline
0.10 & 16/0.67 & 45 & 24/0.63 & 50 & 32/0.67 & 37\\
\hline
0.12 & 24/0.74 &55 & 32/0.67 & 40& 40/0.69 & 12\\
\hline
0.14 & 32/0.77 & 60 & 40/0.70 &37 & 50/0.67 & 12\\
\hline
\end{tabular}
\end{center}
\end{table}

\begin{figure}[hbtp]
\begin{center}
\mbox{\epsfxsize=10cm\epsfysize=8cm\epsffile{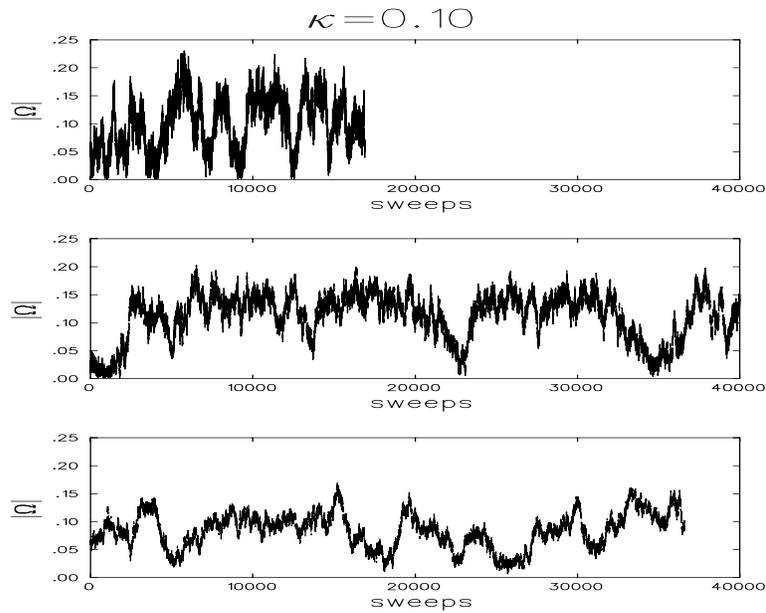}}
\end{center}
\caption{Monte Carlo history of the Polyakov loop for $\kappa=0.10$ on lattices 
$8^3\times 4, \beta=5.670$ (upper),
$12^3\times 4, \beta=5.670$ (middle) and $16^3\times 4, \beta=5.660 $ (lower). }
\label{fig:tunneling_k010}
\end{figure}

\begin{figure}[hbtp]
\begin{center}
\mbox{\epsfxsize=10cm\epsfysize=8cm\epsffile{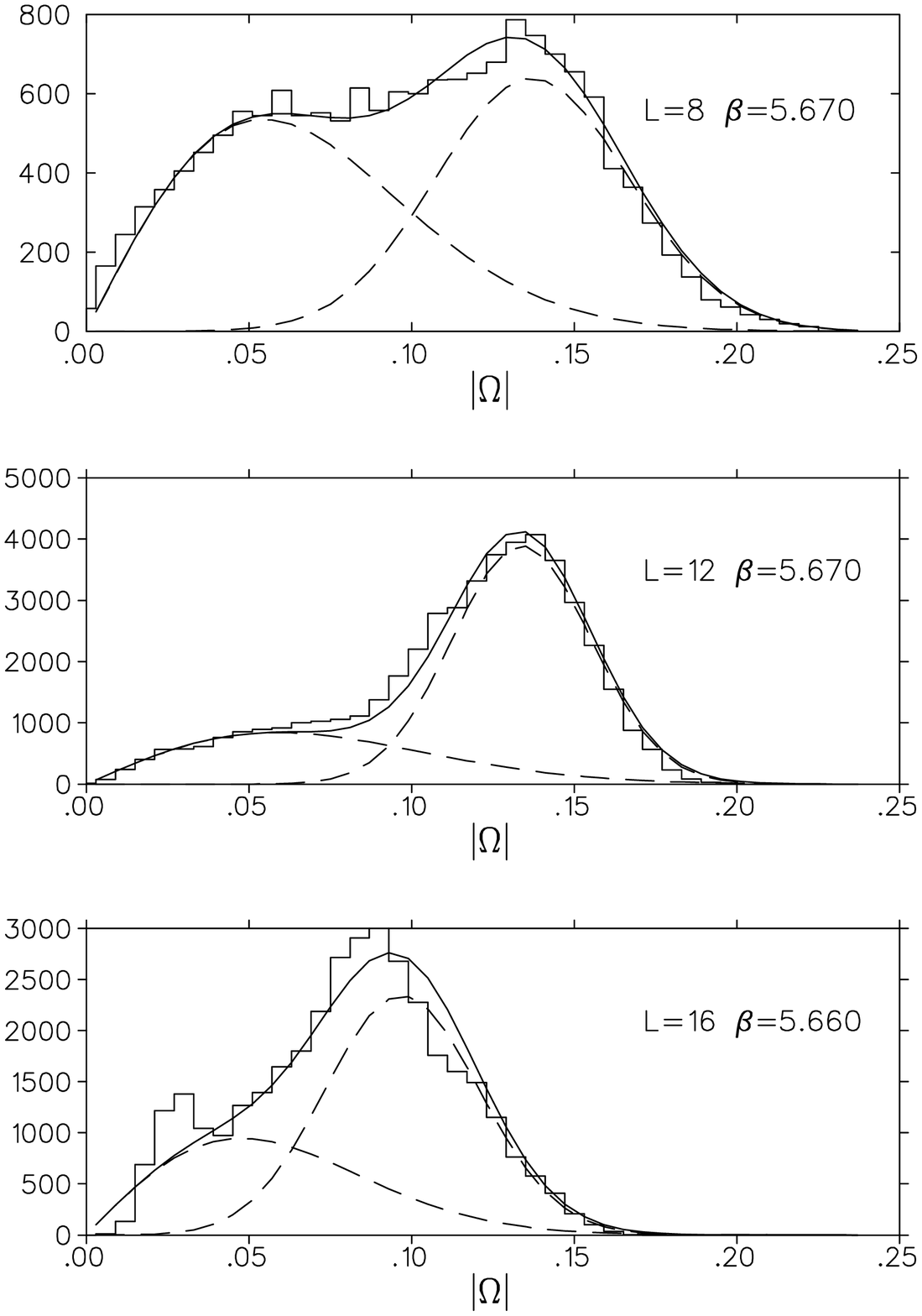}}
\end{center}
\caption{Double peak structure of $|\Omega|$ for $\kappa=0.10$ fitted
to a sum of two gaussians, centered at the origin and at an adjustable
 position on the real axis.} 
\label{fig:polfit}
\end{figure}

\begin{figure}[hbtp]
\begin{center}
\mbox{\epsfxsize=12cm\epsfysize=15cm\epsffile{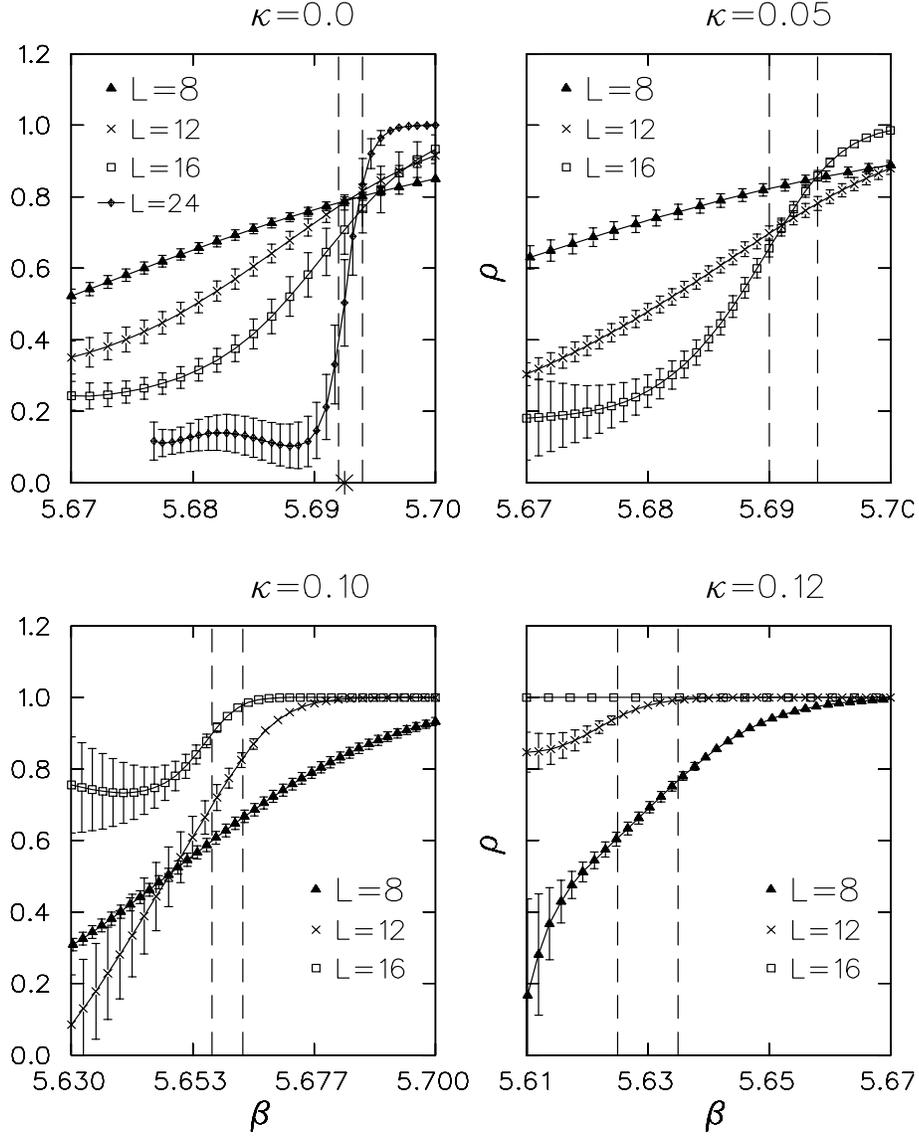}}
\end{center}
\caption{The Deconfinement ratio for $\kappa=0.05$, $\kappa=0.10$ and 
$\kappa=0.12$ for three lattice sizes. The dashed lines give the
error band for  the critical $\beta$ value as determined from
the maximum of the susceptibility (see Fig.~\ref{fig:chi_norm}). The
asterisk denotes the quenched result for $\beta_c$ from ref.~[3]. }
\label{fig:r_deconf}
\end{figure}

\begin{figure}[hbtp]
\begin{center}
\mbox{\epsfxsize=12cm\epsfysize=15cm\epsffile{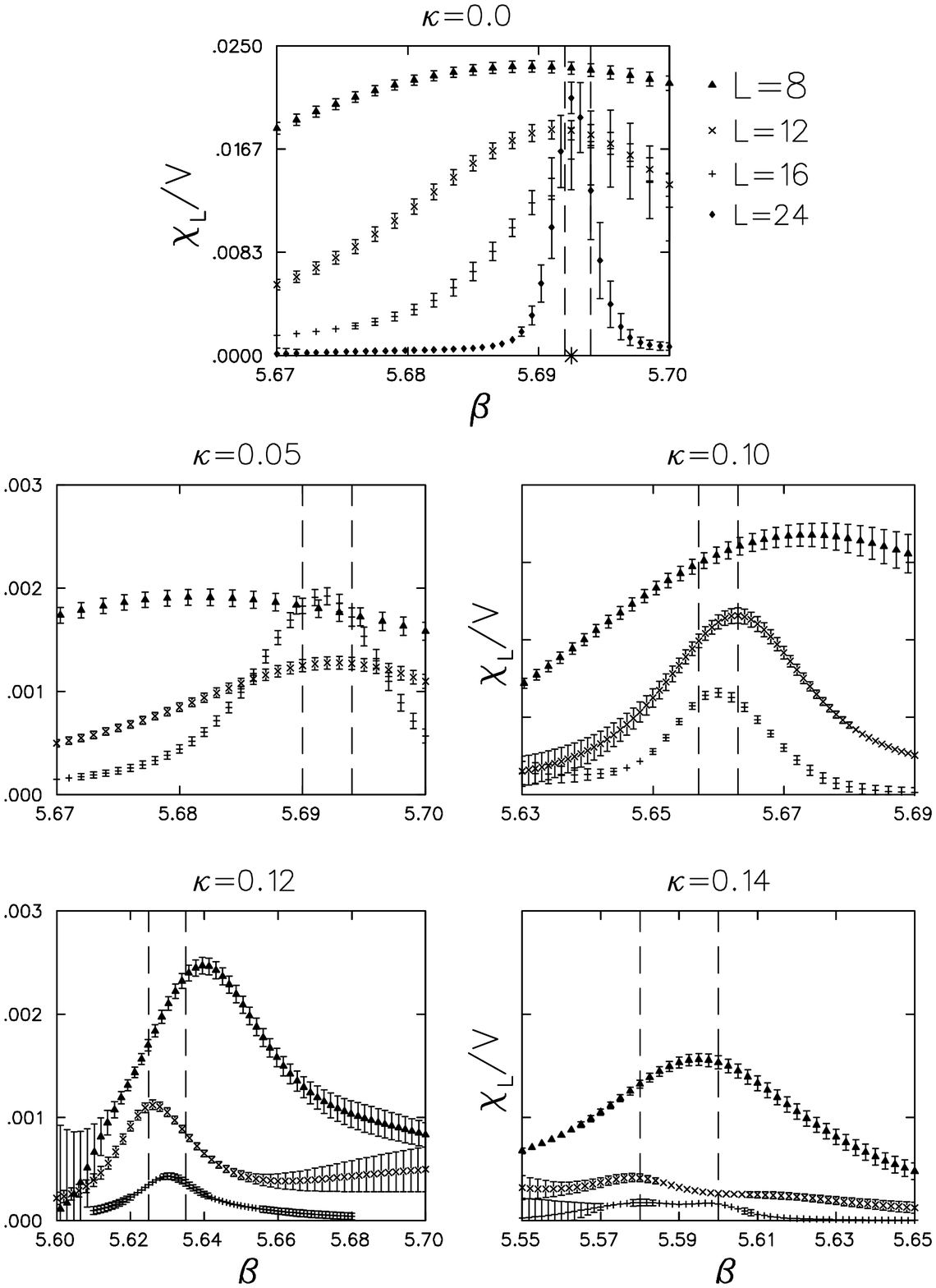}}
\end{center}
\caption{The  susceptibility per unit volume for 
the quenched theory ($\kappa=0$) and for $\kappa=0.05$, $\kappa=0.10$, 
$\kappa=0.12$ and $\kappa=0.14$.  The
asterisk marks  $\beta_c$~[3] for the quenched theory.}
\label{fig:chi_norm}
\end{figure}

Our results for the observables which we used to decide on the order of
the transition are shown 
in Figs.~(\ref{fig:tunneling_k010} - \ref{fig:chi_norm}). 
In Fig. \ref{fig:tunneling_k010}  tunneling  between 
the confined and the deconfined phase is clearly observed 
for $\kappa=0.1$. A similar behaviour is also found   in the case of
 $\kappa=0.05$
whereas for  $\kappa=0.12$ tunneling is no longer observed,
 on our largest, $16^3$, lattice.
The double peak structure of  $|\Omega|$ 
for $\kappa=0.10$ is seen in Fig. \ref{fig:polfit} where
it is fitted to the sum of two gaussian distributions in the complex
plane, one centered at the origin (confined phase) and one centered at
a fitted location along the real axis (deconfined phase).
The second peak seems to approach the first as the lattice size increases,
which would indicate a second rather than a first-order transition.
 A double peak structure was 
 also observed for  $\kappa=0.05$ but  the distance between the peaks
remains the same as the lattice size increased. 
For $\kappa=0.12$ the double peak
was no longer visible.

In Fig.~\ref{fig:r_deconf} we show the deconfinement 
ratio $\rho$ as a function of $\beta$, using
reweighting~\cite{FeSwend},
for $\kappa=0.05, 0.10$ and $0.12$ as well as 
for the quenched case ($\kappa=0$).
The latter is included as a check of our methodology
and statistical accuracy. Our quenched data has similar statistics to our
full QCD data. The intersection of the deconfinement ratio curves $\rho(\beta)$
for various lattice sizes occurs very near the known value of $\beta_c$
\cite{Fuk} shown by an asterisk, and around a value $3/4$ for the deconfinement
ratio. 
The largest lattice ($24^3$) results, actually obtained by reweighting data
from our effective model (see Section 5), help determine the critical point
more accurately, but are not necessary to ascertain the transition itself.
A qualitative change is seen to occur around $\kappa=0.10$, namely 
 the curves for
various lattice sizes stop intersecting. In fact for $\kappa=0.14$, which
is not represented, the deconfinement ratio always stays very near 1.

A similar qualitative change is visible in Fig.~\ref{fig:chi_norm},
which shows the susceptibility $\chi_L$ divided by the volume, again
reweighted as a function of $\beta$.
For $\kappa = 0$ and $0.05$, the peak of $\chi_L / V$ is almost 
independent of $V$. In other words, $\chi_L$ diverges like $V$, as befits
a first-order transition. For all the higher $\kappa$'s, $\chi_L / V$
decreases as the volume grows, indicating that the first-order transition
has disappeared, or that the asymptotic behaviour has not yet set in.

\begin{figure}[hbtp]
\vspace*{-0.5cm}
\begin{center}
\mbox{\epsfxsize=9cm\epsfysize=9cm\epsffile{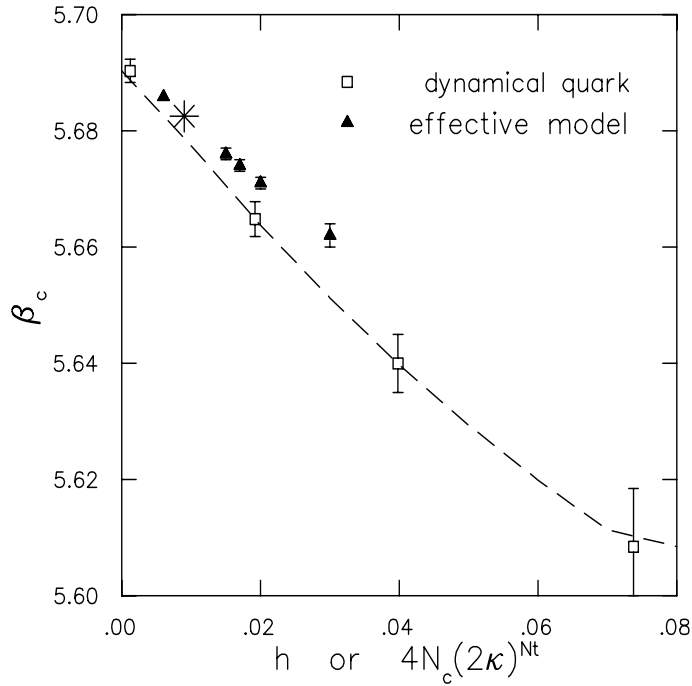}}
\end{center}
\vspace*{-0.5cm}
\caption{The pseudo-critical deconfinement coupling $\beta_c$
as a function of  
$\kappa^{N_t}$ (open squares) found from the analysis of the QCD data 
with one dynamical quark
species (here $N_t=4$). The dashed line  is
to guide the eye. 
The results for the same quantity obtained for our effective model (Section 5)
are also plotted
as a function of the field strength $h$. The asterisk shows the 
second-order end-point. 
The QCD data are shifted upwards by $16N_c \kappa^4$ to take into account 
an effective change in $\beta$ coming from expanding the fermionic determinant
up to order $\kappa^4$.}
\label{fig:results}
\end{figure}

We use the peak of the Polyakov loop susceptibility $\chi_L$ to determine the 
pseudo-critical coupling $\beta_c(\kappa)$. The dotted lines in 
Fig.~\ref{fig:chi_norm} delimitate a band where $\chi_L$ is near its peak
value on our largest volume. 
These bands, also shown in Fig.~\ref{fig:r_deconf},
display the nice agreement between this criterion and the crossing of the
deconfinement ratio lines when that takes place.
We have also looked at the average of the real part of
the Polyakov loop $\Omega$ after projection
to the closest $Z(3)$ axis,
and found that the position where data from different lattices
cross is again within the same error band.
Our values for $\beta_c(\kappa)$
are  collected in Table \ref{table:b_critical}
and plotted in Fig. \ref{fig:results}.
Comparing them with
 available $\beta_c$ values for $N_f=2$~\cite{Aoki}, also included in 
Table~\ref{table:b_critical},
we see that the effect of one dynamical quark on the value of $\beta_c$ is to
shift it from the quenched result of $\beta_c=5.6923(4)$ 
by about half the amount 
of the shift produced by two degenerate quarks for the same $\kappa$ value.   
Not surprisingly, when the dynamical quarks are heavy, their ordering effect
grows linearly with the number of flavours. One could therefore use our $N_f=1$
data to complete the blank $N_f=2$ entries in Table~\ref{table:b_critical}.

\begin{table}[hbtp]
\caption{\label{table:b_critical}}
The pseudo-critical $\beta$ value is given for the 
four $\kappa$ values studied.
The two-flavour results are from ref.~\cite{Aoki} and are given for comparison.
The quenched result from \cite{Fuk} is $\beta_c=5.6923(4)$
\vspace{1ex}
\begin{center}
\begin{tabular}{|c||c|c|c|c|c|}
\hline
&$\kappa$ & 0.05 &0.10 &0.12&0.14  \\ \hline
$N_f=1$ & $\beta_c$ &5.692(2) & 5.660(3)   & 5.630(5) & 5.59(1) \\ \hline
$N_f=2$ &$\beta_c$ & &  & 5.58(2) & 5.46(2) \\ 
\hline
\end{tabular}
\end{center}
\end{table}

\begin{figure}
\begin{minipage}{7cm}
\mbox{\epsfxsize=7cm\epsfysize=7cm\epsffile{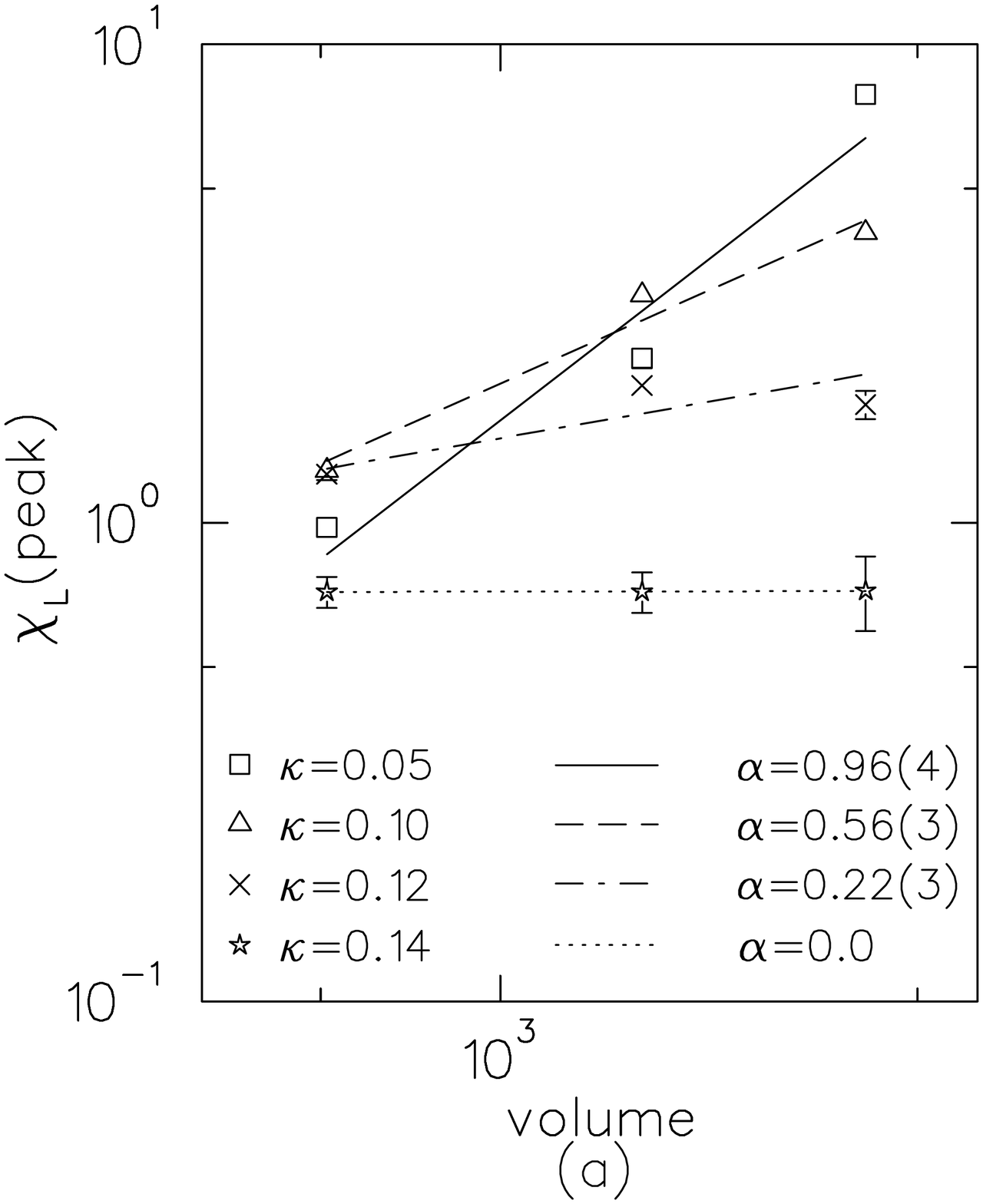}}
\end{minipage} \hfill
\begin{minipage}{7.5cm}
\mbox{\epsfxsize=7cm\epsfysize=7cm\epsffile{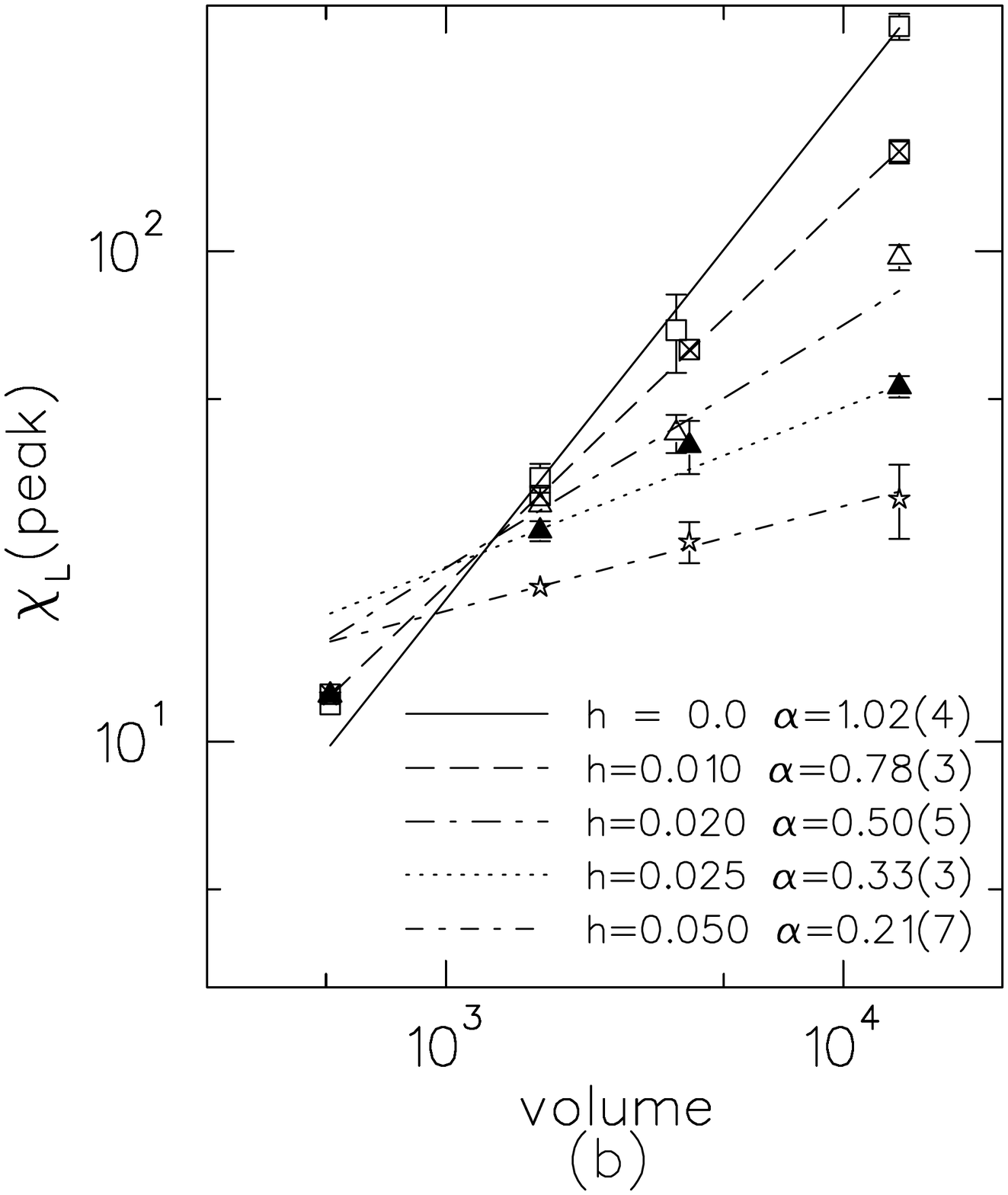}}
\end{minipage} 
\caption{(a) 
The volume dependence of the peak of the susceptibility for $\kappa=0.05$, $\kappa=0.10$, $\kappa=0.12$ and $\kappa=0.14$.
(b) The volume dependence of the peak of the susceptibility 
in our effective model (Section 5)
for $h=0.01$, $h=0.02$, $h=0.025$ and $h=0.05$. A couple of  points are 
slightly shifted in volume for clarity. }
\label{fig:chi_peak}
\end{figure}

The volume dependence of the 
peak value of $\chi_L$ is more clearly displayed 
in Fig.~\ref{fig:chi_peak}a.  
The lines shown are best fits to the form $V^{\alpha}$. 
As discussed in Section 3, values $\alpha = 1$, 
$0 < \alpha < 1$ and $0$ characterize a first-, second-order transition
or crossover respectively, up to finite size corrections.
For $\kappa=0.05$ 
the best fit yields $\alpha=0.96(4)$,
to be compared to the pure gauge value $\alpha=1.02(4)$
for which the data are plotted in Fig.~\ref{fig:chi_peak}b.
The transition is first-order.
For $\kappa=0.14$ on the other hand 
we find $\alpha=0$, a clear signal of crossover behaviour.
For $\kappa=0.10$ and $\kappa=0.12$ the situation is
less clear. The small value of $\alpha=0.22(3)$ at $\kappa=0.12$
favors the crossover region. 
For $\kappa=0.10$, $\alpha=0.56(3)$,
and we must be quite near the second-order end-point.

To summarize our findings:
\begin{itemize}
\item For $\kappa=0.05$ all our observables 
(tunneling, intersecting deconfinement
ratios $\rho(\beta)$, linear dependence of the susceptibility peak on $V$)
behave quite similarly to the quenched case and
consistently point to a first-order transition.
\item For $\kappa=0.12$ and $0.14$, all our observables 
(no tunneling, no crossing of the deconfinement ratios, growth of the
susceptibility peak slow or non-existent)
point to a crossover.
\item
 For $\kappa=0.10$, the situation is less clear. On the largest lattice,
we do not see the signals of a phase transition: 
tunneling is suppressed  as compared to $\kappa=0.05$,
the peak in the Polyakov loop distribution moves towards the origin,
the deconfinement ratio does not intersect the smaller volume curves,
and the peak of the Polyakov loop susceptibility is not much larger than
the peak value for the next smaller volume. 
Based on this largest volume, we consider
$\kappa=0.10$ as being close to the end-point, 
but rather on the crossover side.
\end{itemize}  

A more precise statement would require simulations on larger lattices.
Larger volumes
together with a more refined finite size scaling (FSS)
ansatz would lead to a precise determination of the end point. We carried
out this approach in the context of the effective model discussed 
in detail in the next Section.

\setcounter{equation}{0}
\section{Effective model}

It is clear from the above results that locating the end-point 
$(\beta_{ep},\kappa_{ep})$ accurately requires very large lattice sizes,
beyond our computer capabilities. We thus
adopt a different  approach:
We consider an effective model, in the same universality class as full QCD
but cheaper to simulate. By simulating larger systems, we locate the end-point
in the coupling plane of this model. Then we map this end-point back onto
full QCD.

The simplest prototype of our universality class is the three-dimensional
three-states Potts model in an external field $h$:
\be
S= - \sum_{<nl>} \>\beta \>{\rm Re}\> z^*_n z_l - h\> \sum_n\> {\rm Re}\> z_n
\hspace*{1cm} h > 0
\label{S Potts}
\ee
with $z_n$ an element of $Z(3)$. 
Dimensional reduction at high temperature reduces QCD to this model.
Quenched QCD maps onto the $Z(3)$-symmetric $h=0$ version, which undergoes
a first-order transition at $\beta_c \approx 0.55053$ \cite{Janke}.
Full QCD reduces to the $h > 0$ version, where the strength of the 
$Z(3)$-breaking field $h$ grows with the inverse quark mass.
That model has been studied by DeGrand and DeTar \cite{DeGDeT}.
They found a first-order transition line in the plane $(\beta,h)$, ending
at $(\beta_{ep},h_{ep})$ (see Fig.~\ref{fig:bc_diag}). 
The second-order end-point critical field $h_{ep}$
was evaluated in the mean-field approximation, yielding the small value
$\frac{2}{3}\log2 - \frac{4}{9} \sim 0.018$, and by Monte Carlo where 
the estimate $10^{-3} < h_{ep} < 10^{-2}$ was obtained.
 Unfortunately the mapping from
$h_{ep}$ back to a quark mass is qualitative rather than quantitative.
This is why we have to turn to a more complicated, four-dimensional model.


The starting point is the fermionic determinant which
 can be expanded into loops yielding
\be
{\rm det}({\bf 1} - \kappa M) = exp(- \sum_l \frac{\kappa^l}{l} Tr(M^l))
\label{hopping expansion}
\ee
This expansion will converge quickly for the range of $\kappa$'s under
consideration. The loops which are relevant for the breaking of the $Z(3)$ 
symmetry are the ones that wind around the time direction. Among them the
shortest and most important is the Polyakov loop eq.(\ref{Polyakov loop}), 
which
carries a coefficient 
$\frac{2}{N_t} (2 \kappa)^{N_t}$ 
on a lattice of
time dimension $N_t$. Our effective model tries to incorporate higher-order
effects with an effective $Z(3)$-breaking field $h$, and an action

\be
S_{\rm eff} = S_g[U] -  h(\kappa) \sum_{\bf x} {\rm Re \> Tr} L({\bf x})
\label{S_eff}
\ee

\begin{figure}[hbtp]
\begin{minipage}{8cm}
\mbox{\epsfxsize=8cm\epsfysize=8cm\epsffile{fit_k010_h026.ps2}}
\caption{Best fit of $|\Omega|$ from  full QCD for $\kappa=0.1$ (dashed line) 
to the data obtained from the effective model using reweighting (solid line
for h=0.026).} 
\label{fig:pol match}
\end{minipage} \hfill
\begin{minipage}{7.5cm}
\vspace*{-0.5cm}
\mbox{\epsfxsize=7.5cm\epsfysize=8.8cm\epsffile{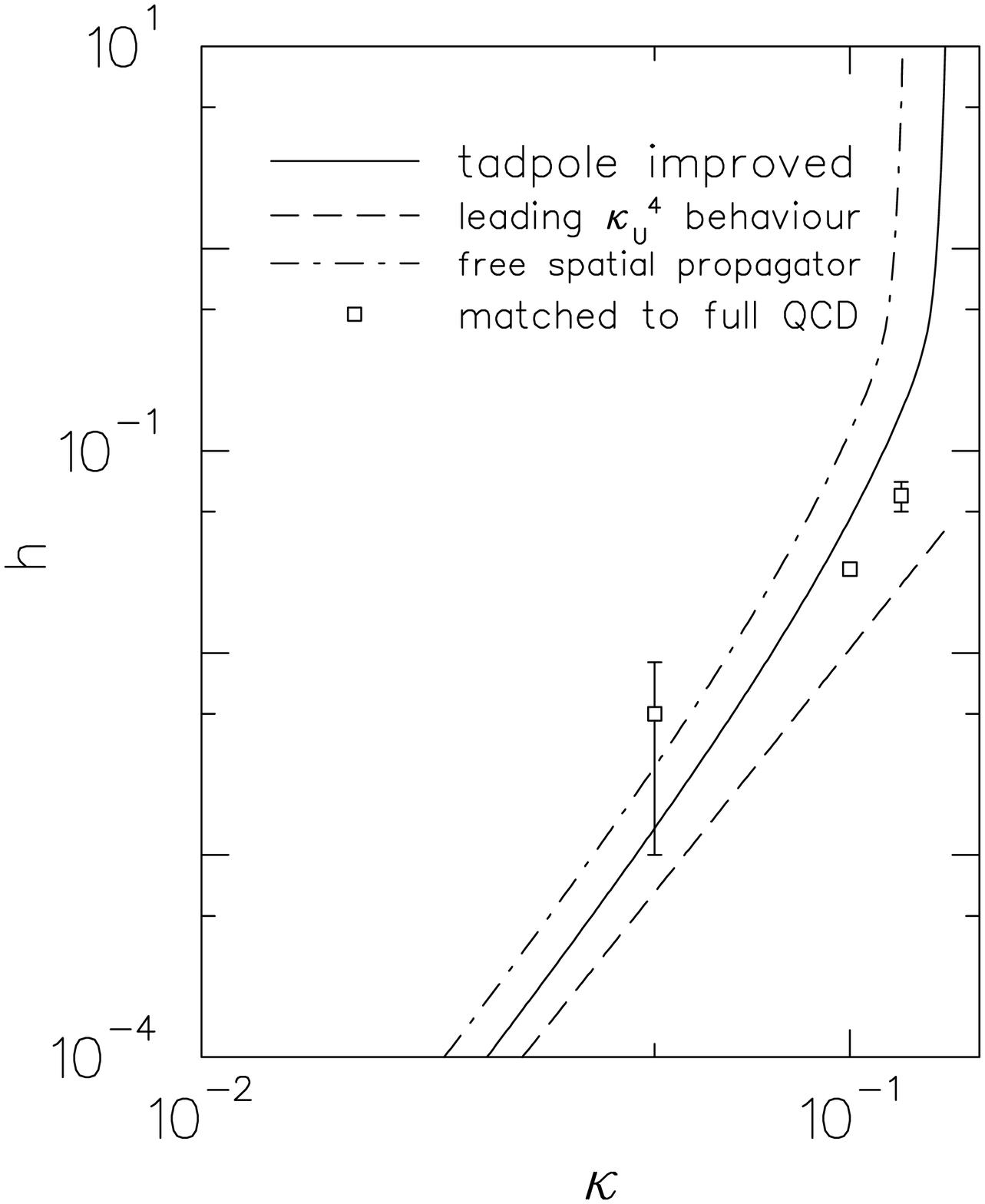}}
\caption{The strength of the $Z(3)$ breaking term
versus $\kappa$}
\label{fig:h parameter}
\end{minipage} 
\end{figure}

Besides the leading hopping parameter expansion above, which should be 
accurate at small $\kappa$, the mapping $h(\kappa)$ can be obtained in
other ways. One way is to resum all spatial hoppings within the free
approximation, so that
\be
h(\kappa) = N_t \int_{-\pi}^{\pi} \> \frac{d^3q}{(2\pi)^3 }
             {\rm Tr} \left [(1+\gamma_0) \> \kappa \> G(q) \>\right]\>^{N_t}
\label{h_perturbative}
\ee
where $G(q)^{-1}$ is the free spatial quark propagator.
This mapping incorporates the divergence of $h$
when $\kappa \rightarrow \kappa_c$, with $\kappa_c = 1/8$. 
This result can be further refined by considering tadpole-improvement,
such that $\kappa$ in (\ref{h_perturbative}) is substituted by
$\kappa \langle plaq \rangle^{1/4}$.
Finally, a non-perturbative approach consists in finding the best match
$h(\kappa)$ between the distribution of the magnitude of the Polyakov loop
$\Omega$ (eq.(\ref{Omega})) in full QCD and that in the
effective model. The quality of such a matching is illustrated in 
Fig.\ref{fig:pol match}. At lower $\kappa$ or $h$, this non-perturbative
matching becomes more sensitive to statistical fluctuations in the 
exploration of $Z(3)$ sectors, which can mimic a symmetry-breaking term.
In that regime the other, analytical methods work better.
Our non-perturbative matching results of the whole $\Omega$ distribution
are listed in Table \ref{table:match}. (Matching the first two moments
of the distribution gave consistent results). Fig.\ref{fig:h parameter}
compares the various mappings $h(\kappa)$ considered here.

\begin{table}
\caption{\label{table:match}}
Non - perturbative determination of  $h$ by fitting to the QCD
data. The value of $h$ listed here gave the best fit  to
 the QCD data of the $|\Omega|$ distribution.\\
\vspace{1ex}
\begin{center}
\begin{tabular}{|c||c|c|c|}
\hline
$\kappa$ & 0.05 &0.10 &0.12 \\ \hline
$h$ & 0.005(4) & 0.026(2)   & 0.06(1)  \\ \hline
\end{tabular}
\end{center}
\end{table}

Now equipped with a reasonably accurate correspondence between $h$ 
and $\kappa$,
we can look for the critical line $(\beta,h)$ and its end-point in our
effective model. We simulate lattices
of spatial size  $8^3, 12^3, 16^3$ and $24^3$ with the temporal direction
fixed at $N_t=4$ as in the case of the full QCD simulations. 
We obtain data at various values of $h=0.01, 0.02, 0.025$ and $h=0.05$
with typically $10 000 - 20 000$ thermalized configurations.

First, we perform the same analysis, using the same observables as 
in Section 4 for full QCD. At each value of $h$, we identify the 
pseudo-critical
coupling $\beta_c(h)$ and obtain the points included in Fig.~\ref{fig:results},
which for small $\kappa$ follows closely the corresponding curve obtained
in full QCD.
In Fig. (\ref{fig:chi_peak}b) we show the dependence of the  peak of 
the Polyakov loop susceptibility $\chi_L$ on the spatial volume  
for various values of $h(\kappa)$. 
The behaviour is again very similar to that of full QCD, indicating a weakening
of the first-order phase transition as $h$ increases, and probably 
a crossover behaviour for $h \geq 0.03$.

\begin{figure}
\begin{minipage}{8cm}
\vspace*{-3cm}
\mbox{\epsfxsize=7cm\epsfysize=7.5cm\epsffile{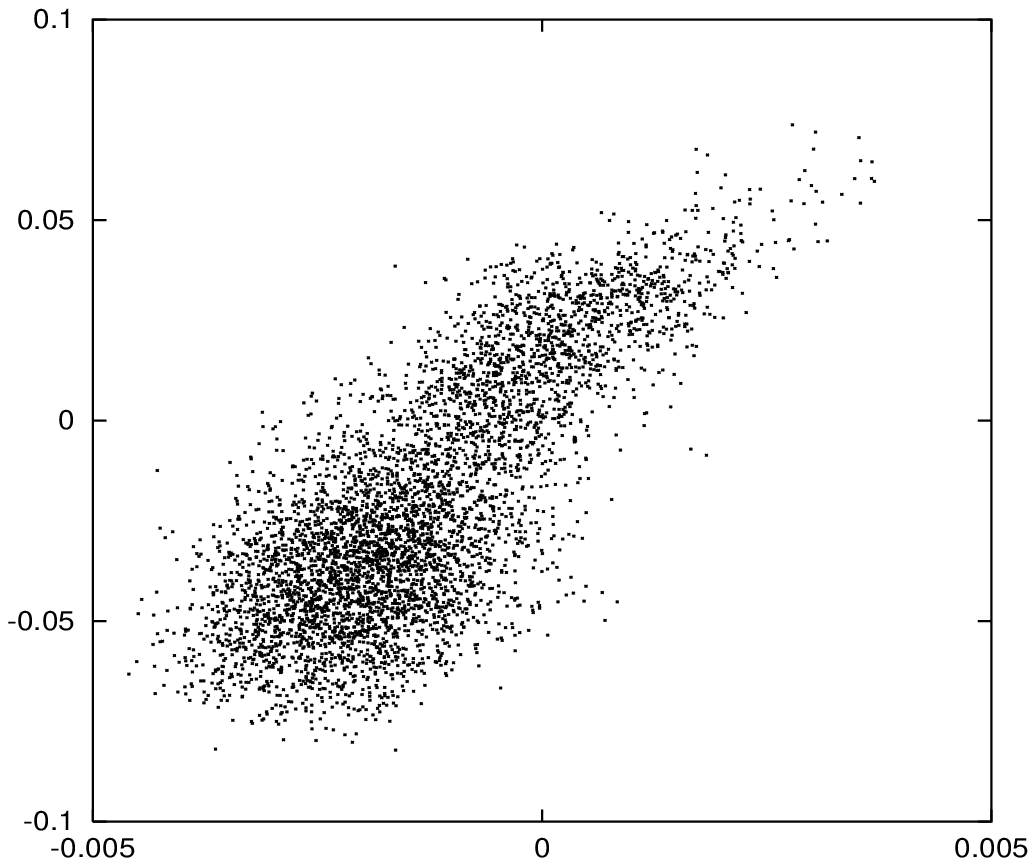}}
\caption{ $(S_{\rm eff}-\langle S_{\rm eff} \rangle)$ vs $(S_g-\langle S_g \rangle)$ 
for 5000 configurations on $24^3$ 
at $h=0.01, \beta=5.680$ }
\label{fig:P_ME}
\end{minipage} \hfill
\begin{minipage}{8cm}
\vspace*{-3cm}
\mbox{\epsfxsize=7cm\epsfysize=7cm\epsffile{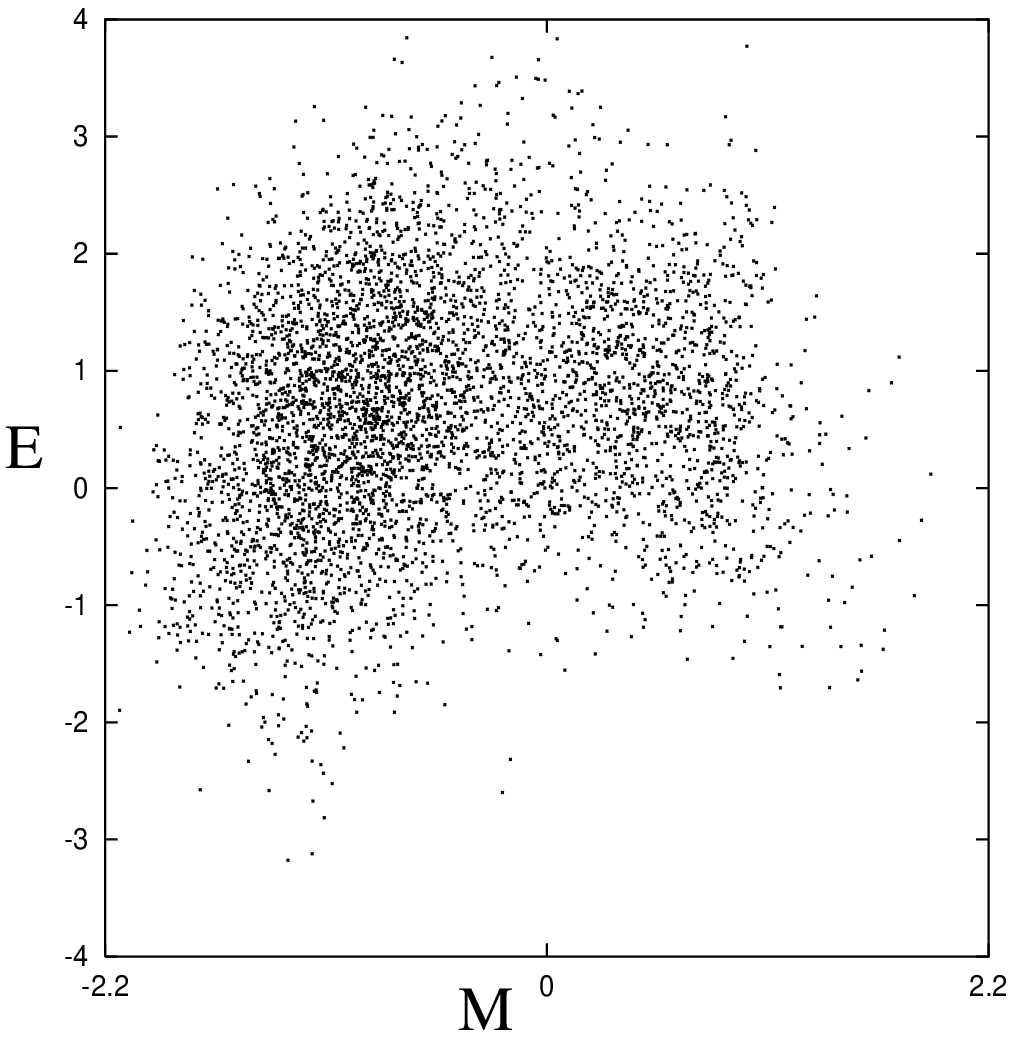}}
\caption{ $(S_{\rm eff}-\langle S_{\rm eff} \rangle)$ vs $(S_g-\langle S_g \rangle)$ for 5000
 configurations on $24^3$ 
at $h=0.01, \beta=5.680$ after a rotation}
\label{fig:P_ME_rot}
\end{minipage} 
\end{figure}

\begin{figure}[hbtp]
\vspace*{-2cm}
\begin{center}
\mbox{\epsfxsize=18cm\epsfysize=14cm\epsffile{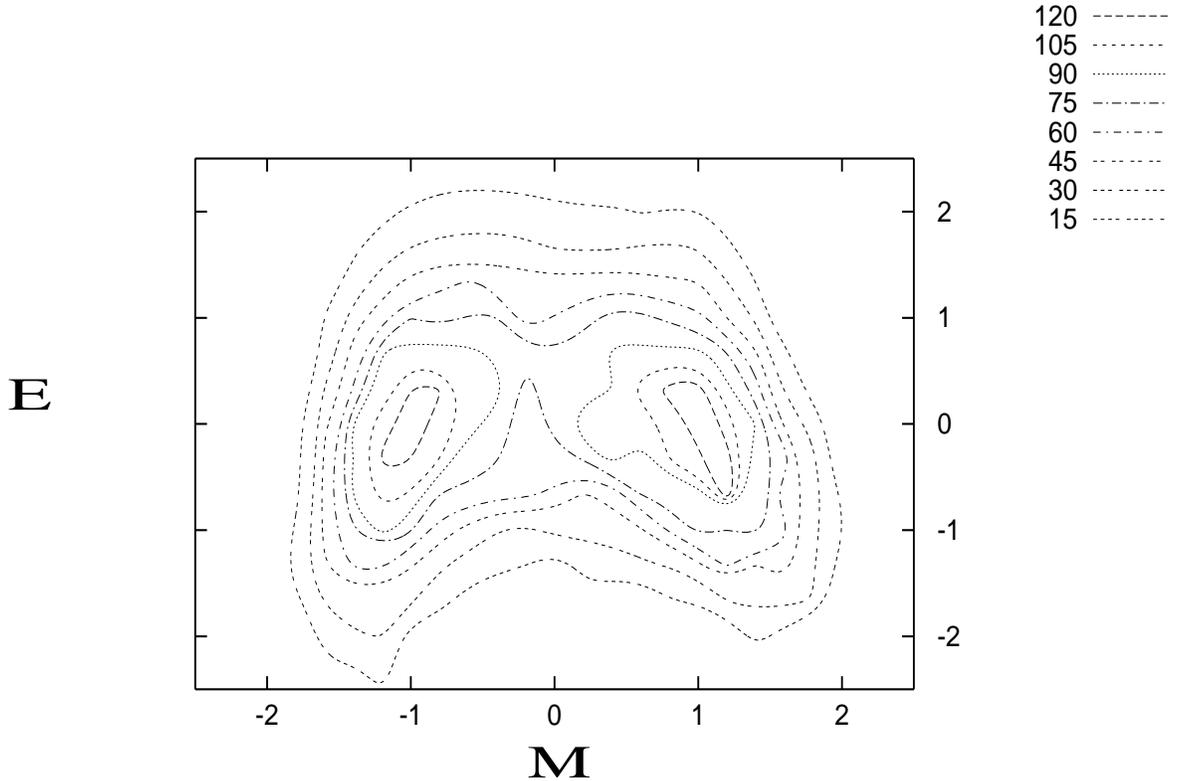}}
\end{center}
\vspace*{-3cm}
\caption{ Normalized probability distribution at the critical point 
$h_{ep}= 0.009$ for $24^3$}
\label{fig:contour}
\end{figure}

Now we would like to determine as precisely as possible the end-point     
$(\beta_{ep},h_{ep})$, and check the scaling behaviour in its vicinity.
In ref.~\cite{EW} the same
issues were addressed in the context of the electroweak theory,
and a reliable numerical procedure was presented which yielded impressively
accurate answers.
The procedure we follow is very similar, although not quite identical.
From our Monte Carlo data we obtain the joint probability 
distribution of the plaquette and the real part of the Polyakov loop 
(shown in fig.~\ref{fig:P_ME}).
The principal axes of this distribution, which diagonalize the correlation
matrix of the two observables, are identified as the magnetization-like
variable $M$ and the energy-like variable $E$. 
(The rotated distribution displayed in fig.~\ref{fig:P_ME_rot} shows 
 the double peak  distribution in the M - like direction.) 
After subtracting the averages from the new variables $M$ and $E$ such that 
$\langle M \rangle = \langle E \rangle = 0$, rescaling them such that
$\langle M^2 \rangle = \langle E^2 \rangle = 1$,
and reweighting in $\beta$ and $h$,
we obtain the probability density $P(M,E)$ shown in Fig.~\ref{fig:contour}
at the end-point. 
We clearly see that the marginal distribution $P(E)$ will show a single peak,
while $P(M)$ will have a double peak. In an infinite volume the distribution
$P(M)$ would be symmetric.
Comparing the  $P(M,E)$ distribution of Fig.~\ref{fig:contour}
 with the $3d$ results  of Fig.~3 Ref.\cite{EW} 
 for the Ising model and the $O(2)$ and $ O(4)$ models, our
result is in closest agreement with
  the  $P(M,E)$ distribution of the Ising model.

\begin{figure}[hbtp]
\begin{minipage}{8cm}
\mbox{\epsfxsize=8cm\epsfysize=10cm\epsffile{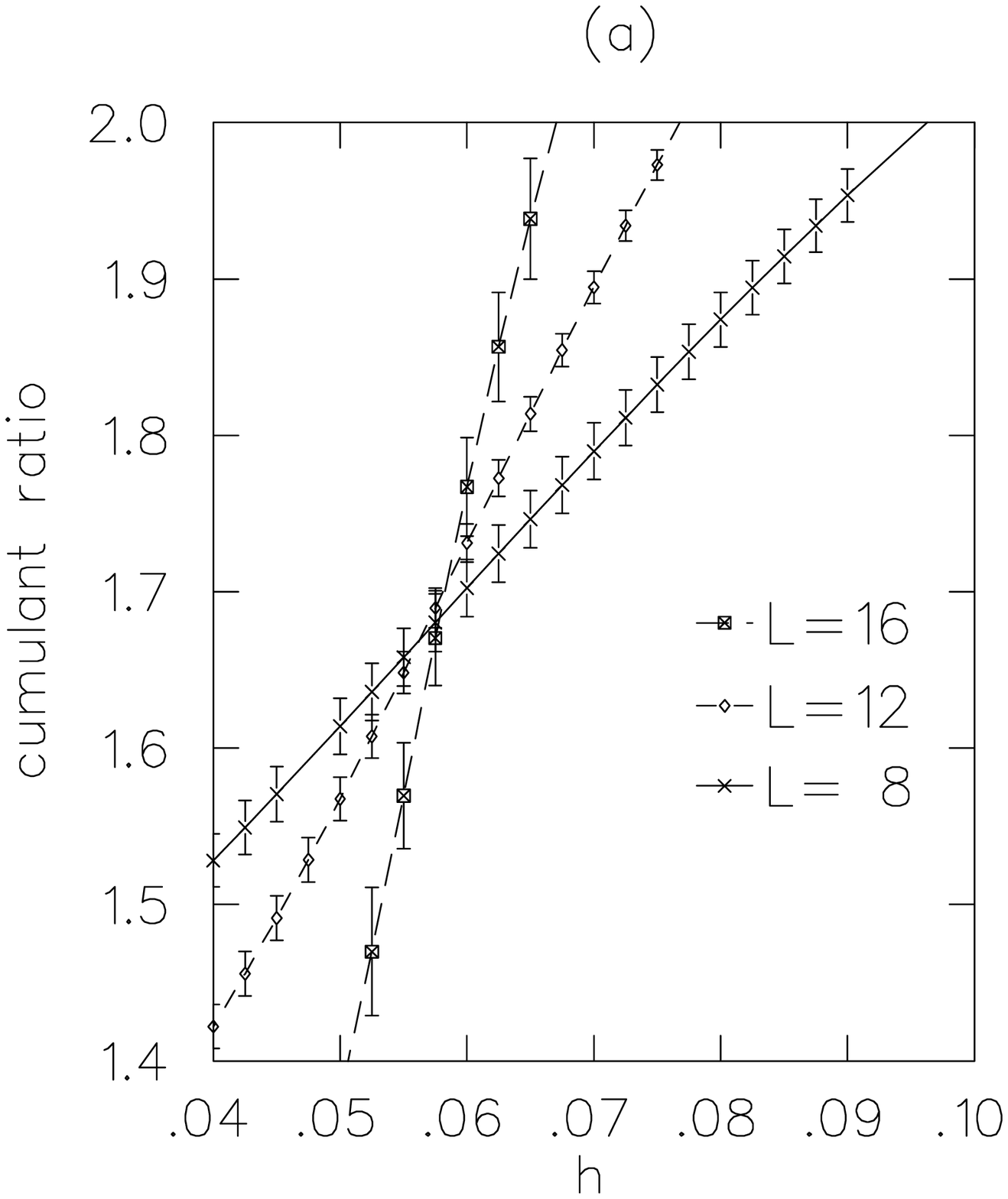}}
\end{minipage} \hfill
\begin{minipage}{8cm}
\mbox{\epsfxsize=8cm\epsfysize=10cm\epsffile{cross_h.ps2}}
\end{minipage} 
\caption{The fourth magnetic cumulant versus the field strength $h$.
(a) for $N_t=2$, 
(b) for $N_t=4$.
The point of intersection of the lines determines 
 the critical point $h_{ep}$. }
\label{fig:crossing}
\end{figure}

\begin{figure}[hbtp]
\begin{center}
\vspace*{-3cm}
\mbox{\epsfxsize=12cm\epsfysize=12cm\epsffile{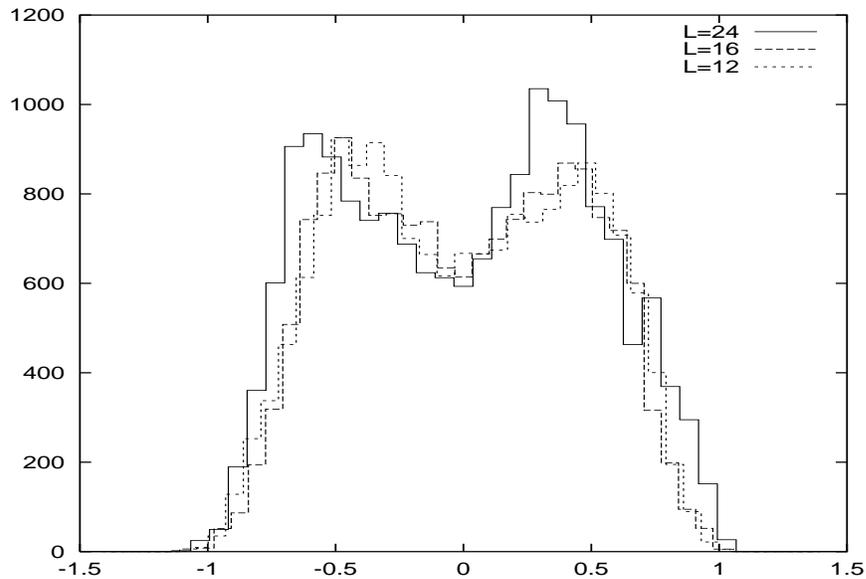}}
\end{center}
\caption{ Scaling of the M - like distribution at the critical point $h_{ep}$
and $\beta_{ep}$ for the three largest lattices. We have applied
a scaling factor of $L^{-0.8}$ to all the data.}
\label{fig:M_scale}
\end{figure}

To determine the end-point we need to perform a finite-size scaling analysis.
For each lattice size, we reweight in $\beta$ and $h$ until we minimize
the asymmetry of the $M$ distribution, by requiring the vanishing of the
third cumulant:
\be
\frac{\langle M^3 \rangle}{\langle M^2 \rangle^{3/2}} = 0 
\label{cumM_3}
\ee
This determines a line in the plane $(\beta,h)$, which must go through
the infinite-volume critical point $(\beta_{ep},h_{ep})$, up to statistical
errors. The intersection of these lines obtained for various lattice
sizes therefore determines the critical point $(\beta_{ep},h_{ep})$.
Parametrizing these lines by the value of $h$, we show in 
Fig.~\ref{fig:crossing} 
the variation of the fourth magnetic cumulant
\be
\frac{\langle M^4 \rangle}{\langle M^2 \rangle^{2}}
\label{cumM_4}
\ee
along these lines, for the three largest lattices considered for $N_t=4$ 
as well
as for three lattices for $N_t=2$. 
The simulation of the  effective model at $N_t=2$
using spatial sizes $8, 12 $ and $16$
can be used as a  check of our procedure in a situation where
significant deviations from scaling are not expected. 
The  three lines for $N_t=2$ cross at a point which   
gives the critical end-point.
The fact that the three lines meet at a point (within statistical
errors) indicates stability against deviations from scaling.
Note  that we made no assumption about the universality class
in our determination of the critical point. Because of its generality,
and because of its relative statistical robustness in our case,
we favor the method we just described to find the end-point over 
the method used in Ref.\cite{EW}, which fixes the fourth cumulant
eq.(\ref{cumM_4}) to its Ising value.
The value of $\langle (\Delta M)^4 \rangle/\langle (\Delta M)^2 \rangle^2 $   
at the critical end-point 
can now be used to check the universality  class of the effective model
and thus the universality class of QCD.
From Fig.~\ref{fig:crossing}(a) at $h_{ep}=0.06$ 
we find $\langle (\Delta M)^4 \rangle/\langle (\Delta M)^2 \rangle^2 \sim 1.67(5) $
in good agreement with
 the corresponding value of $1.604(1)$  for the Ising 
model\footnote{For the $O(2)$ and $O(4)$ models
 $\langle (\Delta M)^4 \rangle/\langle (\Delta M)^2 \rangle^2=1.233(6)$~\cite{Hasen} and 
$1.092(3)$~\cite{KK} respectively 
i.e. clearly lower than our value of $1.67(5)$.}.
 The value of $\beta_{ep}$ determined from the crossing of the
lines parametrized with the value of $\beta$ is $\beta_{ep}=5.047(1)$.
For $N_t=4$  
the lines for our three largest volume converge at
 $h \sim 0.009$. Scaling violations are still seen for the lattice with
spatial extent $L=12$. Thus here 
the end - point is  determined less accurately as compared to the
 $N_t=2$ case, although we have increased our statistics five fold using 
 $\sim 100,000$ configurations for the two smaller lattices and $\sim 50,000$  
for the largest lattice. Since the point of intersection
for the two largest lattices is within statistical errors  
consistent with $h=0.009$ we take $h_{ep} \sim 0.009(1)$
as the best determination of the critical point from these data.
Having determined the critical point we can look at critical
exponents. The magnetic susceptibility at $h_{ep}$ 
should scale like~\cite{EW} 
\be
\chi_M = V\langle (\Delta M)^2 \rangle \propto L^{-3+\gamma/\nu} 
\label{chi_M}
\ee
We  plot the  $\chi_M$ distributions in 
Fig.~\ref{fig:M_scale} for our three  lattices after  scaling 
 with $L^{-0.8}$. They are seen to nicely fall on top of each other. 
This scaling thus yields a rough estimate of the  exponent
$\gamma/\nu \sim 2.2$ to be compared with
a value of $1.96$ for the Ising and  $O(2),\> O(4)$ model. 
To determine exponents like $\alpha/\nu$ which can pin down the 
Ising model universality class
we need to
consider the scaling behaviour of the E - like variable, which
is smaller that the M - like variable by at least two orders of magnitude.
Scaling based on the E - like variable was found to be
unstable within the statistical precision of our present simulation.

Finally, we comment on the Polyakov distribution
displayed in Fig.~\ref{fig:contour}.
It is clear from this figure that even at the end-point $h_{ep}=0.009$, 
the Polyakov loop distribution still shows a marked double peak structure.
In ref.~\cite{MOgil} where the same effective model was studied,
the criterion used to identify the end-point was to look
for the value of $h$ where the double peak 
structure of the real part of the Polyakov loop was no longer visible.
This criterion thus led to a much bigger value $\sim 0.08$ 
for the critical field strength than we are finding here.

\newpage

\section{Discussion}
 
From the previous section we conclude that the end-point of the first-order
transition occurs at $h(\kappa) \sim 0.009$ which maps to $\kappa \sim 0.08$.
The qualitative picture expected for (2+1) flavours  in the
continuum is that there is a finite region in the mass plane $(m_{u,d},m_s)$
of the two degenerate flavours versus the third near the quenched limit 
where the 
deconfinement phase transition is first order. 
Our $N_f=1$ result corresponds to the infinite mass limit for $m_{u,d}$.
In this limit, which up to now received little attention, we find  
that the first order pure gauge phase transition persists
up to $\kappa \sim 0.08$.  
Since we expect dynamical quark effects to be twice as strong for two flavours 
(see Table~\ref{table:b_critical})
we can estimate the boundary of the first-order region as
\be
	2 h_{eff}(k_{u,d}) + h_{eff}(k_s) \approx 0.009
\ee
with $h_{eff}(\kappa)$ as per Fig.\ref{fig:h parameter}.

It is of course of crucial importance to establish the scaling behaviour
of our results, and to express the end-point parameter as a quark mass,
not a hopping parameter.
One might wonder if the rather small value of $\kappa_{ep}$ we found may not 
be an artifact of the relatively coarse lattice discretization, and if one
would recover $\kappa_{ep}=0$ in the continuum limit. This would mean
that the deconfinement transition disappears as soon as dynamical quarks
of any mass are allowed. Such a scenario goes against the robustness 
of a first-order phase transition. The first-order deconfinement transition
must be robust against small variations of the parameters, before ending
at a second-order point. We made this argument in the introduction to
justify the expected phase diagram Fig.\ref{fig:bc_diag} in our discrete
theory. The same argument applies to the continuum QCD theory.
Therefore, we expect the deconfinement transition to end at a finite
quark mass $m_q$. 
Determining this mass would require a full-blown scaling study of 1-flavour
QCD, which is beyond the scope of this paper.
However, we have considered what happens to our effective model as the
lattice spacing $a$ changes.
We have simulated our effective model for $N_t=2$ and found
$h_{ep} \approx 0.06$, with scaling exponents consistent with those 
at $N_t=4$. 
Changing $N_t=4$ to $N_t=2$ amounts to doubling the lattice spacing $a$.
Therefore, the variation of the end-point value $h_{ep}$ indicates a
scaling behaviour
\be
h_{ep}(2a) / h_{ep}(a) \approx 0.06 / 0.009 \propto a^{2.7}
\ee
This strongly supports the existence of a continuum limit
for the effective model eq.(\ref{S_eff}), with action
\be
S_{eff} = S[A] - \tilde{h} \int d^3x L({\bf x})
\ee
where $h = \tilde h a^3$ in the discretized theory, having a critical
end-point at some positive $\tilde h_{ep}$.

If one believes in the physical nature of the critical coupling 
$\tilde h_{ep}$, then the corresponding quark mass cannot be infinite.
For instance, the leading hopping parameter expansion 
eq.~(\ref{hopping expansion}) 
at constant physical temperature $T=(N_t a)^{-1}$ gives 
\be 
h(\kappa) \sim 4 N_c (2 \kappa)^{(aT)^{-1}}
\ee
If $h(\kappa) = \tilde h_{ep} a^3$, then 
this relation is not consistent with 
$lim_{a \rightarrow 0}\kappa = 0$. One is again lead to expect
persistence of the first order transition, for some range of quark mass,
in the continuum one-flavour QCD theory.

We can get an estimate of the critical quark mass in physical units
by using the naive prescription 
$m_q a = \frac{1}{2} (\frac{1}{\kappa} - \frac{1}{\kappa_c})$.
Using the quenched data of ref.~\cite{Wein}
for the pion mass we find $\kappa_{c}=0.1694(2)$, at $\beta=5.7$ in the
zero-flavour theory. On the other hand the SESAM
collaboration finds 
$\kappa_{c}=0.1585(1)$~\cite{SESAM} at $\beta=5.6$ in the two-flavour theory. 
Going to our value of  $\beta_{ep}=5.683(3)$ will decrease the critical
value from SESAM slightly. 
If we neglect this change and take the average between the
zero- and two-flavour cases, we end up with $\kappa_{c} \approx 0.164$.    
Taking the end-point value $\kappa_{ep} \sim 0.08$ 
 we find a quark mass  $m_q a =3.2$ in units of the lattice spacing $a$.
This mass is in fact of the same order as that of the doubler modes,
which leads to  an overestimation of the mass at the end point.
Since $m_q a > 1$,
finite $a$ corrections are expected to
be rather large. At tree level  the corrected mass  is 
given by
\be
m_q a  \sim \frac{m_q a}{\sqrt{1 + m_q a}}
\label{treelevelmass}
\ee
which reduces our estimate by about a factor of two to  $m_q a \sim 1.56$.
To convert this  to physical units  
we use $(4 a)^{-1} \sim 220 MeV$ from the deconfinement 
temperature. This gives $m_q \sim 1.4 GeV$ for the end-point.


A critical quark mass of the order of a  GeV is in line 
with phenomenological expectations. 
The pure gauge deconfinement
transition is fairly weak, with a critical correlation length 
$\xi_c \sim (7-10) a$ at $\beta=6$~\cite{Aoki} 
i.e. $\sim 11 GeV^{-1}$ or
$\cal O$(5 $\sigma^{-1/2}$). 
This is the minimum system size necessary
to observe the first-order nature of the deconfinement transition.
Dynamical quarks introduce a new length
scale $r_c$, namely the distance where the string breaks, 
$r_c \sim {\cal O}(2 m_q /\sigma)$. Confinement can only be observed up to
this distance. 
For very heavy quarks, $r_c > \xi_c$: string breaking occurs for a large 
separation, larger than the critical correlation length. 
The passage from confinement to deconfinement will allow ``liberated'' quarks
to appear even if their separation is less than $r_c$. This qualitative
change signals a phase transition.
As the quark mass is lowered, the string-breaking scale $r_c$ decreases.
When $r_c \sim \xi_c$, the passage from confinement to deconfinement
does not liberate quarks that were not already liberated by string breaking.
There is no qualitative change from one regime to the other, and
one cannot really tell if the system is confined or deconfined: the
phase transition has disappeared and been replaced by a crossover.
This occurs for
\be
m_q \sim {\cal O}(5 \sqrt\sigma / 2) 
\hspace*{1cm} i.e.\hspace*{1cm}  m_q \sim {\cal O}(1) GeV
\label{end point mass}
\ee

\setcounter{equation}{0}

\section{Summary and Conclusions}
The multiboson method can be used to simulate an odd number of flavours 
as well as an even number.
In this work we have shown that the multibosonic algorithm
 is well suited
for the study of one - flavour QCD for  moderately 
heavy Wilson quarks. 
Using this algorithm  we were able to
carry out a detailed  Finite Size Scaling (FSS) analysis to determine the  
pseudocritical  $\beta(\kappa)$ line for $\kappa$ values  up to $\kappa=0.14$. 
We  demonstrated
that the first
order phase transition seen in the quenched theory persists when one includes 
dynamical quarks. Using FSS  we 
showed that the deconfinement phase transition 
gets weaker as the dynamical quark
 mass increases, then turns into a crossover. 
In general we find that the
dynamical quark effects on the phase transition are approximately half those 
for two flavours.
Using lattices of sizes $8^3, 12^3$ and $16^3 \times 4$
within full QCD, we found an end - point around $\kappa=0.1$.
A more accurate determination would require simulation of 
larger spatial volumes.
In order to carry out 
a more refined FSS analysis with larger lattices
 we  considered an effective model in the same 
universality class as full  QCD. 
In this effective model the effects of the fermionic determinant 
were simulated by an effective $Z(3)$ - breaking  field coupled to 
the Polyakov loop.
We studied the phase transition as a function of this field strength $h$
on lattices with spatial volumes $8^3, 12^3, 16^3 $ and $24^3$ and
found a first order transition that gets weaker as the field strength
increases, exactly like in QCD as $\kappa$ increases. 
Performing a FSS analysis using the
joint probability distribution of the plaquette and the real part of the
Polyakov loop we were able to determine, without any assumptions about the
universality class of the model,
 the end point of the first order transition line.
Matching of the Polyakov loop histograms or its 
first two moments in the effective model to those in full QCD
enabled us to determine  non - perturbatively  
the correspondence between $h$ and $\kappa$.
In this way we obtained the
value  of $\kappa_{ep}\sim 0.08$ for the end - point of the QCD first order 
transition line.
Futhermore, the FSS analysis of the effective model at the critical point 
yielded results consistent 
with our effective model (and one-flavour QCD) being in the same 
universality class as the Ising model.

We have also studied our effective model 
on a coarser lattice at $N_t=2$, and have found good scaling behaviour
of the end-point field strength, with similar Ising-like exponents at
the end-point. 
This scaling test strongly supports the existence of 
a critical end-point at a non-vanishing positive field strength
in the continuum limit of our effective model. 
In turn, this indicates a non-vanishing quark mass for the end-point
of the deconfinement transition line $T_c(m_q)$ in continuum one-flavour QCD.
Our findings are in agreement with the idea that a first-order 
transition is robust against small variations of the parameters, before
ending at a second-order point.
  Converting the value of $\kappa_{ep}=0.08$  to physical units can only
be done approximately, using 
quenched and two-flavour determinations of $\kappa_c$
at similar $\beta$ values.
After correcting
for finite lattice spacing errors 
to tree level we obtain an estimate of
the quark mass at the end - point of
$m_q \sim 1.4 $~GeV, consistent with phenomenological expectations.
A simulation of one-flavour QCD closer to the continuum limit
would allow tighter control over discretization errors, but is
beyond our computer resources.

\vspace*{2cm}

{\bf Acknowledgements:} We thank SIC of the EPFL in
Lausanne, ZIB in Berlin and  the Minnesota Supercomputing Institute
for providing the necessary computer time.
Ph. de F. would like to thank M. Ogilvie, K. Rummukainen, A.J. van der Sijs and 
J. Zinn-Justin, and C.A. B. Svetitsky for discussions.

\end{document}